\documentclass[twocolumn,showpacs,preprintnumbers,amsmath,amssymb,]{revtex4}
\usepackage{graphicx}% Include figure files
\usepackage{dcolumn}% Align table columns on decimal point
\usepackage{bm}% bold math
\usepackage{subfigure}
\def\b{\begin{equation}}
\def\e{\end{equation}}
\def\le{\langle}
\def\re{\rangle}
\begin{document}
\title{Relativistic quantum nonlocality for the three-qubit\\ Greenberger-Horne-Zeilinger state}% Force line breaks with \\
\author{Shahpoor Moradi}
 \altaffiliation[]{}%Lines break automatically or can be forced with \\
 \email{shahpoor.moradi@gmail.com}
\affiliation{%
Department of Physics, Razi University, Kermanshah, Iran
}%
\date{\today}% It is always \today, today,
\begin{abstract}
Lorentz transformation of three-qubit Greenberger-Horne-Zeilinger
(GHZ) state is studied. Also we obtain the relativistic spin joint
measurement for the transformed state. Using these results it is
shown that Bell's inequality is maximally violated for three-qubit
GHZ state in relativistic regime. For ultrarelativistic particles
we obtain the critical value for boost speed which Bell's
inequality is not violated for velocities smaller than this value.
 We also show that in ultrarelativistic limit Bell's
inequality is maximally violated for GHZ state.
\end{abstract}

  \maketitle

Relativistic effects on quantum entanglement and quantum
information is investigated by many authors [1-13]. Alsing and
Milburn  \cite{am} studied the Lorentz  transformation of
maximally entangled states. By explicit calculation of the Wigner
rotation they described the observation of the entangled Bell
states from two inertial frames moving with the constance velocity
with respect to each other. They concluded that entanglement is
Lorentz invariant. Terashima, \textit{et al}. \cite{tu1}
considered to relativistic Einstein-Podolsky-Rosen correlation and
Bell's inequality. They showed that the degree of the violation of
Bell's inequality decreases with increasing the velocity of the
observers if the directions of the measurement are fixed. They
extended these considerations to the massless case. Ahn,
\textit{et al}. \cite{ahn1} investigated the Bell observable for
entangled states in the rest frame seen by the moving observer and
showed that the entangled states satisfy the Bell's inequality
when the boost speed approaches the speed of light. D. Lee,
\textit{et al}. \cite{lee} showed that maximal violation of the
Bell's inequality can be achieved by properly adjusting the
directions of the spin measurement even in a relativistically
moving inertial frame. Kim, \textit{et al}. \cite{wt} obtained an
observer-independent Bell's inequality, so that it is maximally
violated as long as it is violated maximally in the rest frame.
They showed that the Bell observable and Bell states for Bell's
inequality should be transformed following the principle of
relativistic covariance, which results in a frame independent
Bell's inequality.

In this paper we would like to study the Bell's inequality for
three-qubit GHZ state in relativistic regime. For doing this, we
need the Lorentz transformation of GHZ state and relativistic spin
joint measurement.

 The following paper is organized as follows. First
 we review the representation of the Lorentz group and
Wigner little group. Then we calculate the Lorentz transformation
of three-qubit GHZ state. After that  we obtain the relativistic
spin joint measurement of GHZ state and calculate the degree of
violation for a special case which in non relativistic case gives
the maximally violation of Bell's inequality. Finally we calculate
the degree of violation for GHZ state when particles moving with
same momentum and particles moving in the center of mass frame.
Finally we compare our results with two qubit case.

 A multipartite state is expressed by \b
\Phi_{\vec{p}_1\sigma_1,\vec{p}_2\sigma_2,...}=a^{\dag}(\vec{p}_1,\sigma_1)
a^{\dag}(\vec{p}_2,\sigma_2)...\Phi_0,\e where $\vec{p}_i$ is
three momentum vector, $\sigma_i$ is spin label, $a^{\dag}$ is
creation operator and $\Phi_0$ is Lorentz invariant vacuum state.
Multipartite state (1) has the Lorentz transformation property
\cite{wien} $$
U(\Lambda)\Phi_{\vec{p}_1\sigma_1,\vec{p}_2\sigma_2,...}=$$\b
\sum_{\bar{\sigma}_1\bar{\sigma}_2...}D^{(j_1)}_{\bar{\sigma}_1\sigma_1}\left(W(\Lambda,p_1)\right)
D^{(j_2)}_{\bar{\sigma}_2\sigma_2}\left(W(\Lambda,p_2)\right)...\Phi_{\vec{p}_{1\Lambda}\sigma_1,\vec{p}_{2\Lambda}\sigma_2,...}.\e
Here $\vec{p}_{1\Lambda}$ is the three vector part of ${\Lambda
p}_1$, $D^{(j)}_{\bar{\sigma}{\sigma}}$ is the unitary spin-$j$
representation of the three dimensional rotation group, and
$W(\Lambda,p)$ is Wigner's little group element \b
W(\Lambda,p)=L^{-1}(\Lambda p)\Lambda L(p), \e where  $L(p)$ is
the standard boost that takes a particle of mass $m$ from rest to
four-momentum $p^{\mu}$. Transformation of creation operator is \b
U(\Lambda) a^{\dag}\left(\vec{p},\sigma\right)U^{-1}(\Lambda)
=\sum_{\sigma'}D^{(j)}_{\sigma'\sigma}(W(\Lambda,p))a^{\dag}(\vec{p}_{\Lambda},\sigma')
.\e The Wigner representation of the Lorentz group for
spin-$\frac{1}{2}$ is \b
D(W(\Lambda,p))=\cos{\frac{\delta_{\vec{p}}}{2}}+
i(\vec{\sigma}\cdot{\vec{n}})\sin{\frac{\delta_{\vec{p}}}{2}}=\left(%
\begin{array}{cc}
  D_{00} & D_{01} \\
  D_{10} & D_{11} \\
\end{array}%
\right),\e with\b
\cot{\frac{\delta_{\vec{p}}}{2}}=\coth\frac{\xi}{2}\coth\frac{\chi}{2}+\hat{e}\cdot\hat{p},
\e where
\b\cosh{\chi}=(p^0/m),\;\;\;\;\;\;\tanh{\xi}=\beta=v/c,\;\;\;\;\;\;\vec{n}=\hat{e}\times\hat{p}.\e
Here $\hat{e}$ is a normal vector in the boost direction and $v$
is the boost speed. We consider the case in which the boost speed
is perpendicular to momentums of particles. In this case we have\b
\cos{\frac{\delta}{2}}=\left[
\frac{(1+\sqrt{1-\beta^2})(\cosh{\chi}+1)}{2(\sqrt{1-\beta^2}+\cosh{\chi})}
\right]^{1/2},\e \b \sin{\frac{\delta}{2}}=\left[
\frac{(1-\sqrt{1-\beta^2})(\cosh{\chi}-1)}{2(\sqrt{1-\beta^2}+\cosh{\chi})}
\right]^{1/2}, \e where in ultrarelativistic limit as
$\beta\rightarrow1$ take the forms \b
\cos{\frac{\delta}{2}}\rightarrow\left[
\frac{1+\textrm{sech}{\chi}}{2} \right]^{1/2}, \e \b
\sin{\frac{\delta}{2}}\rightarrow\left[
\frac{1-\textrm{sech}{\chi}}{2} \right]^{1/2}. \e

Investigations show that exist a family of pure entangled $N>2$
qubit states that do not violate any Bell's inequality for
N-particle correlations for the case of a standard Bell experiment
on N qubits \cite{gis}. For $N=3$, one class is
Greenberger-Horne-Zeilinger (GHZ) state given by
$|GHZ\re=\frac{1}{\sqrt{2}}(|000\re+|111\re)$, the other class is
represented by the W state
$|W\re=\frac{1}{\sqrt{3}}(|110\re+|101\re+|011\re)$, where 0 and 1
represent spins polarized "up" and "down" along the z axis.

We can express  GHZ  state using creation operator in the rest
frame
 $$
|GHZ\re=\frac{1}{\sqrt{2}}\left\{a^{\dag}(\vec{p}_1,0)a^{\dag}(\vec{p}_2,0)a^{\dag}(\vec{p}_3,0)\right.$$\b\left.+
a^{\dag}(\vec{p}_1,1)a^{\dag}(\vec{p}_2,1)a^{\dag}(\vec{p}_3,1)\right\}\Phi_0.\e
For simplicity we assume that momentum of particles are
sufficiently localized around momentum $\vec{p}_i$. Realistic
situation involve the wave pockets with gaussian form
$\exp(-{\vec{p}_i}^2/2{\Delta}^2)$ with characteristic spread
$\Delta$. Note that these particles are indistinguishable. Authors
in reference [12] investigated  that one can create
distinguishable qubits from indistinguishable particles by
preparing particles in minimum uncertainty states that are well
localized with a sharp momentum. They show that N-qubit product
state can be constructed from N single particle states as \b
|\psi\re_N=\otimes_{n=1}^Ne^{-iaP_z}|\psi\re_1, \e where
$|\psi\re_1$ is a single particle state. State (13) describes a
one dimensional latices of particles with separation $a$. Using a
proton (hydrogen atom) in the millikelvin range as an example,
condition for distinguishablity is $a\gg 100 A^{\circ}.$

Using relation $(4)$ Lorentz transformation of GHZ state becomes
\[
|GHZ'\re=\frac{1}{\sqrt{2}}(A|000\re +B|001\re +C|010\re +D|011\re
\]\b+E|100\re +F|101\re +G|110\re +H|111\re)|\vec{p}_1\vec{p}_2\vec{p}_3\re_{\Lambda},\e
with
\[
A=D^1_{00}D^2_{00}D^3_{00}+D^1_{01}D^2_{01}D^3_{01},\]
\[B=D^1_{00}D^2_{00}D^3_{10}+D^1_{01}D^2_{01}D^3_{11}
,\]
\[
C=D^1_{00}D^2_{10}D^3_{00}+D^1_{01}D^2_{11}D^3_{01},\]
\[D=D^1_{00}D^2_{10}D^3_{10}+D^1_{01}D^2_{11}D^3_{11}
,\]
\[
E=D^1_{10}D^2_{00}D^3_{00}+D^1_{11}D^2_{01}D^3_{01}\]
\[F=D^1_{10}D^2_{00}D^3_{10}+D^1_{11}D^2_{01}D^3_{11}
,\]
\[
G=D^1_{10}D^2_{10}D^3_{00}+D^1_{11}D^2_{11}D^3_{01},\] \b
H=D^1_{10}D^2_{10}D^3_{10}+D^1_{11}D^2_{11}D^3_{11}, \e where
$D^i$ is Wigner representation for particle $i$. The
generalization of the Bell's type inequality to the case of three
particles is the one proposed by Mermin which can be expressed in
terms of correlation functions as follows \cite{mer} \b
\varepsilon=|E(\vec{a},\vec{b},\vec{c'})+E(\vec{a},\vec{b'},\vec{c})+
E(\vec{a'},\vec{b},\vec{c})-E(\vec{a'},\vec{b'},\vec{c'})|\leq 2,
\e where$$
E(\vec{a},\vec{b},\vec{c})=\le\psi|(\vec{a}\cdot\vec{\sigma})
\otimes(\vec{b}\cdot\vec{\sigma})\otimes(\vec{c}\cdot\vec{\sigma})|\psi\re,
$$is correlator function, $\vec{a}$, $\vec{b}$ and $\vec{c}$ are
real three-dimensional vectors of unit length and
$\vec{\sigma}=(\sigma_x,\sigma_y,\sigma_z)$ is the Pauli spin
operator. For each  measurement, one of two possible alternative
measurement is performed: $\vec{a}$ or $\vec{a'}$ for particle 1,
$\vec{b}$ or $\vec{b'}$ for particle 2, $\vec{c}$ or $\vec{c'}$
for particle 3. For  GHZ state, Bell's inequality is violated if,
for example, measurements are made in the xy plane along some
appropriate directions. In this case \b
E(\vec{a},\vec{b},\vec{c})=\cos(\phi_1+\phi_2+\phi_3),\e where we
labelled the angles from the x-axis. The correlation function
$E(\vec{a},\vec{b},\vec{c})$ can take the value either +1 or -1
under both realistic theory and quantum mechanical theory, thus
the maximum value of $\varepsilon$ is 4.

 The normalized
relativistic spin observable $\hat{a}$ is given by \cite{cz}
\b\hat{a}=\frac{(\sqrt{1-\beta^2}\vec{a}_\bot+\vec{a}_\|)\cdot
\vec{\sigma}}{\sqrt{1+\beta^2[(\vec{e}\cdot\vec{a})^2-1]}},\e
where the subscripts $\bot$ and $\|$ denote the components which
are perpendicular and parallel to the boost direction. Operator
$\hat{a}$ is related to the Pauli-Lubanski pseudo vector which is
relativistic invariant operator corresponding to spin.
 Now we are ready to
calculate the relativistic Bell's inequality for three particles
system. Spin joint measurement for the transformed  state
$|GHZ'\re$  for measurement in xy plane is 
\[
{\le GHZ}'|\hat{a}\otimes \hat{b}\otimes \hat{c}|GHZ'\re\]\[=
\left\{[1+\beta^2(a_x^2-1)][1+\beta^2(b_x^2-1)][1+\beta^2(c_x^2-1)]\right\}^{-1/2}\]
\[\times\Re(
E^*Da_{xy}b^*_{xy}c^*_{xy}+F^*Ca_{xy}b^*_{xy}c_{xy}
\]
\b +G^*B a_{xy}b_{xy}c^*_{xy}+H^*Aa_{xy}b_{xy}c_{xy}),\e where
$a_{xy}=a_x+ia_y\sqrt{1-\beta^2} $ and so on. In ultra
relativistic limit as $\beta\rightarrow1$ we get
\[
\le GHZ'|\hat{a}\otimes \hat{b}\otimes
\hat{c}|GHZ'\re\]\b\rightarrow
\frac{a_xb_xc_x}{|a_xb_xc_x|}\Re\{AH^*+G^*B+F^*C+E^*D\}, \e which
is not correlated. In non-relativistic limit
\[
\le GHZ'|\hat{a}\otimes \hat{b}\otimes
\hat{c}|GHZ'\re\]\[\rightarrow
a_xb_xc_x-a_yb_xc_y-a_yb_yc_x-a_xb_yc_y\]\b
=\cos(\phi_1+\phi_2+\phi_3). \e
 Here we consider to
the vector set inducing the maximal violation of Bell's inequality
for GHZ state in non relativistic case. With the following
suitably chosen measurement settings,
\[
\vec{a}=\vec{b}=\vec{c}=\hat{y},
\]
\b\vec{a'}=\vec{b'}= \vec{c'}=\hat{x}, \e and using the algebra of
pauli matrices we have \b
(\sigma_x\sigma_x\sigma_x-\sigma_y\sigma_y\sigma_x-\sigma_y\sigma_x\sigma_y-\sigma_x\sigma_y\sigma_y)|GHZ\re=4|GHZ\re,
\e
 then for GHZ state Bell's
inequality is maximally violated with $|\varepsilon|=4$. For set
vectors $(22)$ the relativistic Bell measurement  becomes \b
\varepsilon'=4\Re\left\{AH^*\right\}. \e We obtain the degree of
violation for two cases.

\textit{Cass I}. $\vec{p}_1=\vec{p}_2=\vec{p}_3=p\hat{z}$

 In this case
\[
D(W(\Lambda,p_1))=D(W(\Lambda,p_2))=D(W(\Lambda,p_3))\]\b=\left(\begin{array}{cc}
 \cos{\frac{\delta}{2}} & -\sin{\frac{\delta}{2}} \\
  \sin{\frac{\delta}{2}}& \cos{\frac{\delta}{2}}\\
\end{array}\right),
\e and Bell observable takes the form
\b\varepsilon'=\cos^3{\delta}+3\cos{\delta}.\e In
ultrarelativistic limit as $\beta\rightarrow1$, (26) reduces to \b
\varepsilon' \rightarrow \textrm{sech}^3{\chi}+3
\textrm{sech}{\chi}\leq 4.\e In this limit amount of violation for
very high energy particles goes to zero, but for low energy
particles approaches to 4, similar to nonrelativistic limit
$\beta\rightarrow 0$.

 \textit{Cass II}. $\vec{p}_1+\vec{p}_2+\vec{p}_3=0$

 The particles are
in the center of mass frame with the following momentums
 \[\vec{p}_1=-p\hat{z},\]\[\vec{p}_2=\left(\frac{1}{2}\hat{z}+\frac{\sqrt{3}}{2}\hat{y}\right)p
 ,\]\b
\vec{p}_3=\left(\frac{1}{2}\hat{z}-\frac{\sqrt{3}}{2}\hat{y}\right)p.\e

 Wigner representations of the the Lorentz group for particles
1, 2 and 3 respectively are written as \b
D(W(\Lambda,p_1))=\left(\begin{array}{cc}
 \cos{\frac{\delta}{2}} & \sin{\frac{\delta}{2}} \\
  -\sin{\frac{\delta}{2}}& \cos{\frac{\delta}{2}}\\
\end{array}\right)
,\e\[
D(W(\Lambda,p_2))=D^*(W(\Lambda,p_3))\]\b=\left(\begin{array}{cc}
 \cos{\frac{\delta}{2}}+i\frac{\sqrt{3}}{2}\sin{\frac{\delta}{2}} & -\frac{1}{2}\sin{\frac{\delta}{2}} \\
  \frac{1}{2}\sin{\frac{\delta}{2}}& \cos{\frac{\delta}{2}}-i\frac{\sqrt{3}}{2}\sin{\frac{\delta}{2}}
\\
\end{array}\right).
\e In this case Bell observable to be
\b\varepsilon'=\frac{1}{16}\cos^3\delta+\frac{3}{8}\cos^2\delta+\frac{33}{16}\cos\delta+\frac{3}{2}.\e
which for ultrarelativistic limit as $\beta\rightarrow1$  reduces
to
\b\varepsilon'\rightarrow\frac{1}{16}\textrm{sech}^3\chi+\frac{3}{8}\textrm{sech}^2\chi+\frac{33}{16}\textrm{sech}\chi+\frac{3}{2}.\e
For very high energy particles amount of violation is
$\varepsilon'=1.5$, but for low energy particles $\varepsilon'=4$
which is maximally violation of Bell's inequality.

From the two preceding cases it is obvious that, the degree of
violation decreases under Lorentz transformation. This is because
Bell observable is evaluated with the same spin measurement
directions as in the non-relativistic lab frame. By finding a new
set of spin measurement directions, for example by rotating the
spin measurement directions with Wigner rotation, Bell's
inequality is still maximally violated in a Lorentz-boosted frame
\cite{tu1,lee,wt}.

 It's interesting to express Bell
observable in order function of $\beta$ for  ultrarelativistic
particles. In this situation relations $(8)$ and $(9)$ reduce to
\b \cos{\frac{\delta}{2}}\approx\left[
\frac{(1+\sqrt{1-\beta^2})}{2} \right]^{1/2},\e \b
\sin{\frac{\delta}{2}}\approx\left[ \frac{(1-\sqrt{1-\beta^2})}{2}
\right]^{1/2},\e  then the amount of violation (26) takes the form
\b \varepsilon'\approx\sqrt{1-\beta^2}(4-\beta^2).\e It's obvious
that critical value $\beta_c$ for satisfying Bell's inequality  is
$0.8$. Critical value for case II is $0.97$.

 Now we compare our
results with two-qubit case obtained by Ahn, \textit{et al}
\cite{ahn1}. They calculated relativistic Bell observable for two
qubit entangled Bell state, when particles move in the center of
mass frame, and found \b
\varepsilon'=\frac{2}{\sqrt{2-\beta^2}}(\sqrt{1-\beta^2}+\cos2\delta).\e
In ultrarelativistic limit $\beta\rightarrow1$:
$\varepsilon'\rightarrow|4\textrm{sech}^2\chi-2|\leq2$ which
indicates the Bell's inequality is not violated in this limit.
This result is not same as three-qubit case. For very high energy
particles (36) reduces to \b
\varepsilon'\approx\frac{2}{\sqrt{2-\beta^2}}(1+\sqrt{1-\beta^2}-2\beta^2),\e
the critical value for violation of Bell's inequality in this case
is $\beta_c=0.86$, which is smaller than three-qubit case when
particles move in the center of mass frame.

 In conclusion using
Bell's inequality, we studied the nonlocal quantum properties of
GHZ state in relativistic formalism. First we obtained the
relativistic spin joint measurements for Lorentz transformed
three-qubit GHZ state. We show that in ultrarelativistic limit
joint measurement is uncorrelated. We also investigated the degree
of violation for particles moving with same momentum and particles
moving in the center of mass frame. Bell's inequality is maximally
violated in rest frame or in moving frame with rest particles, but
as  seen by moving observer is not always violated, because the
degree of violation of Bell's inequality depends on the velocity
of the particles and observer. In non relativistic case the spin
degrees of freedom and momentum degrees of freedom are
independent. But in relativistic regime Lorentz transformation of
spin of particle depends on its momentum. For GHZ state we show
that in ultrarelativistic limit Bell's inequality is maximally
violated which is not same as two-qubit case. Finally, for very
high energy particles we obtained a critical value for satisfying
Bell's inequality. The critical value for three-qubit state  is
greater than two-qubit case.
\subsection*{ACKNOWLEDGMENTS} It is a pleasure to thank Professor E. Solano  for his valuable suggestions.

\end{document}